\documentclass[%
 reprint,
superscriptaddress,
 amsmath,amssymb,
 aps,
prx,
]{revtex4-2}

\usepackage{graphicx}
\usepackage{dcolumn}
\usepackage{amsmath,amssymb,amsthm,mathtools}
\usepackage{bm}
\usepackage{physics}
\usepackage{siunitx}
\usepackage{braket}
\usepackage{xcolor}
\usepackage{enumitem}
\usepackage{amsthm}

\newtheorem{lemma}{Lemma}  
\newtheorem{theorem}{Theorem}

\providecommand{\me}{\mathrm{e}}

 \providecommand{\mi}{\mathrm{i}}
 \providecommand{\mS}{\mathrm{S}}
 \providecommand{\mE}{\mathrm{E}}

\begin{document}

\preprint{APS/123-QED}

\title{Exact Markovian Dissipation Requires Singular Energy Resources}

\author{Hiroki Nakabayashi}
 \email{nakaba@iis.u-tokyo.ac.jp}
\affiliation{Department of Physics, The University of Tokyo, 5-1-5 Kashiwanoha, Kashiwa, Chiba 277-8574, Japan}
\affiliation{Analytical Quantum Complexity RIKEN Hakubi Research Team, RIKEN Center for Quantum Computing (RQC), 2-1 Hirosawa, Wako, Saitama 351-0198, Japan}


\date{\today}

\begin{abstract}
The Gorini--Kossakowski--Lindblad--Sudarshan (GKLS) equation describes irreversible quantum dynamical semigroups. We show that this description cannot be exact under physically regular energy conditions. We prove that the open-system survival probability under physically regular energy conditions has sublinear decay, whereas any dissipative GKLS semigroup has a linear short-time decay. Hence exact Markovian dissipation requires singular energy resources: an unbounded-below total Hamiltonian or infinite initial energy, and a divergent interaction-energy moment. Therefore, a dissipative time-independent GKLS equation should be regarded as an effective description rather than the exact reduced dynamics of a Hamiltonian dilation satisfying physically regular energy conditions.
\end{abstract}

\maketitle


\section{Introduction}
The Gorini--Kossakowski--Lindblad--Sudarshan (GKLS) \cite{gorini1976completely,lindblad1976generators} equation is one of the most widely used effective descriptions of irreversible quantum dynamics, with applications in quantum optics \cite{gardiner2004quantum,jager2022lindblad,link2020dynamical}, condensed matter physics \cite{prosen2008third,karevski2013exact,ekman2024liouvillian}, quantum information \cite{verstraete2009quantum,schwartzman2025modeling,malekakhlagh2025efficient}, and many-body physics \cite{nathan2020universal,mori2024liouvillian,aydougan2025stabilizing}. 
It is characterized by the semigroup law: the dynamics is generated by a time-independent
generator and describes memoryless relaxation and decoherence.

However, derivations of GKLS equations from the total Hamiltonian dynamics typically rely on Born--Markov and secular approximations \cite{breuer2002theory,rivas2012open}. 
A more basic question is whether these approximations are merely technical conveniences,
or whether exact Markovian dissipation itself requires an idealization that cannot be
realized as the reduced dynamics of a physically regular Hamiltonian dilation.

The Chiu--Sudarshan--Misra (CSM) theorem showed that, for a Hamiltonian bounded from below and an initial state with finite energy expectation, the survival probability cannot exhibit exponential decay at short times \cite{chiu1977time}.
However, this theorem concerns the survival probability $\abs{\braket{\psi(0)|\psi(t)}}^2$ of a closed system, with $\ket{\psi(t)}$ evolved from $\ket{\psi(0)}$, and does not address reduced open-system dynamics.

In this paper, we prove an open-system counterpart of the CSM short-time constraint. For a finite-dimensional system, we consider the reduced survival probability
\begin{gather}
    p(t)=\Tr_{\rm S}[\rho_{\rm S}(0)\rho_{\rm S}(t)].
\end{gather}
We show that, for every pure initial state, the decay of this survival probability is
sublinear near $t= 0$,
\begin{gather}
    1-p(t)=o(t),
\end{gather}
provided that either of the two independent regularity assumptions on the Hamiltonian dilation holds.
The first is a global assumption: the total Hamiltonian is bounded from below and has finite
expectation in every product state of an arbitrary pure state of the system of interest and the fixed
environmental state. This is the direct open-system analogue of the CSM energy condition.
The second is a local interaction condition: given a microscopic decomposition
\(H=H_0+H_{\rm int}\), the second moment of the interaction Hamiltonian evolved under the
free dynamics remains finite at short times.

This short-time structure is incompatible with any genuinely dissipative time-independent GKLS semigroup. 
For such a semigroup, the same reduced survival probability satisfies
\begin{gather}
1-p(t)=\alpha t+\mathcal{O}(t^2),~~\alpha> 0,
\end{gather}
for at least one pure initial state, whenever the dissipative part of the GKLS generator is non-zero. 
Thus, an exact non-unitary GKLS semigroup must violate both regularity conditions established here. This conclusion explains why deriving a time-independent dissipative GKLS equation from regular microscopic unitary dynamics requires a Markovian approximation or a singular limiting procedure.

The issue is not whether a unitary dilation exists mathematically, but whether it can be generated by a physically regular Hamiltonian.
Stinespring's theorem \cite{stinespring1955positive} ensures that every completely positive,
trace-preserving map admits a unitary dilation on an enlarged Hilbert space, and Davies showed
that finite-dimensional quantum dynamical semigroups can be represented as reductions of
strongly continuous unitary groups \cite{davies1972some}.
These results, however, do not guarantee that the dilation is generated by a physically regular Hamiltonian. Recently, it has been shown that non-unitary finite-dimensional semigroups admit no bounded
time-independent Hamiltonian dilations \cite{vom2023quantum}. However, typical open-system
environments, such as bosonic baths, are described by unbounded Hamiltonians
\cite{leggett1987dynamics,hahn2025efficiency}. Our result rules out a broader physically relevant class of Hamiltonian dilations even in
the unbounded case. Exact Markovian dissipation should therefore be understood as a singular energy-resource
limit, rather than as the reduced dynamics of a regular Hamiltonian system.

\section{Main result: a short-time conflict}

We now present the main result. The conflict can be summarized as
follows: a dissipative GKLS semigroup leads to a linear decay at $t=0$,
whereas any time-independent Hamiltonian dilation with regular energy
conditions gives a sublinear decay. Thus, an exact dissipative GKLS semigroup cannot be realized under regular energy conditions.

Hereafter, we define $\mathcal{H}_{\rm S}$ and $\mathcal{H}_{\rm E}$ as
the Hilbert spaces of the system and the environment, respectively, and
$I_{\rm S}$ and $I_{\rm E}$ as the identity operators on these spaces.
Let the system be initially in a pure state
$\rho_{\rm S}=|\psi\rangle\langle\psi|$, and define
\begin{gather}
\rho_0(0)
:=
|\psi\rangle\langle\psi|\otimes \rho_{\rm E},
~~
P
:=
|\psi\rangle\langle\psi|\otimes I_{\rm E},
~~
Q:=I-P
\end{gather}
with fixed environmental state $\rho_{\rm E}$, where $I$ is the identity operator on the total system.
For an exact reduced dynamics generated by a time-independent total
Hamiltonian $H$,
\begin{gather}
\rho_{\rm S}(t)
=
\Tr_{\rm E}
\left[
\me^{-\mi Ht}\rho_0(0) \me^{\mi Ht}
\right].
\end{gather}
We quantify the survival loss from the initial system
state by
\begin{align}
\ell(t)
&:=1-
\Tr_{\rm S}
\left[
\rho_{\rm S}(0)\rho_{\rm S}(t)
\right]
\\
&=
1-
\Tr_{\rm S}
\left[
|\psi\rangle\langle\psi|\rho_{\rm S}(t)
\right]
\\
&=
\Tr
\left[
Q \me^{-\mi Ht}\rho_0(0) \me^{\mi Ht}
\right].
\end{align}

We now show that a linear loss is forbidden by two complementary
finite-resource assumptions: one global and one interaction-based.
Note that, because the conditions below involve possibly unbounded positive
self-adjoint operators, $\Tr[\rho A]$ is understood for such an
operator $A$ as the squared Hilbert--Schmidt norm
$\norm{A^{1/2}\rho^{1/2}}_2^2\in[0,\infty]$.

\begin{theorem}[Global regularity]
\label{thm:spectral-regularity}
Let $\mathcal{H}_{\rm S}$ be finite-dimensional, and let the reduced
dynamics be generated by a time-independent total Hamiltonian $H$ as
\begin{gather}
\rho_{\rm S}(t)
=
\Tr_{\rm E}
\left[
\me^{-\mi Ht}
(\rho_{\rm S}\otimes\rho_{\rm E})
\me^{\mi Ht}
\right],
\end{gather}
with a fixed environmental state $\rho_{\rm E}$ and a self-adjoint
Hamiltonian $H$ on
$\mathcal{H}_{\rm S}\otimes\mathcal{H}_{\rm E}$.
Suppose that the total Hamiltonian is bounded from below, $H\geq E_g I$,
and
\begin{gather}
\Tr
\left[
\rho_0(0) (H-E_g I)
\right]
<\infty
\end{gather}
for every pure state $|\psi\rangle$.
Then, for every pure state $|\psi\rangle$,
\begin{gather}
\lim_{t\to +0}
\frac{\ell(t)}{t}
=
0 .
\end{gather}
\end{theorem}

This condition means that the total Hamiltonian has a lower bound and a
finite expectation value with respect to the pure state of the system of interest and the
fixed initial environmental state.

Theorem 1 is the open-system counterpart of the
Chiu-Sudarshan-Misra analysis of unstable quantum states: a survival
probability $|\braket{\phi(0)|\phi(t)}|^2$ for a closed system
$\ket{\phi(t)}=\me^{-\mi Ht}\ket{\phi(0)}$ cannot decay linearly at short
times if the Hamiltonian is bounded from below and the initial state has
finite energy $\braket{\phi(0)|H|\phi(0)}<\infty$
\cite{chiu1977time}. Here the same mechanism constrains the
reduced survival probability of an open system rather than the survival
probability of a closed unstable state.

\begin{theorem}[Interaction regularity]
\label{thm:interaction-regularity}
Let $\mathcal{H}_{\rm S}$ be finite-dimensional, and let the reduced
dynamics be generated by a time-independent total Hamiltonian $H$ as
\begin{gather}
\rho_{\rm S}(t)
=
\Tr_{\rm E}
\left[
\me^{-\mi Ht}
(\rho_{\rm S}\otimes\rho_{\rm E})
\me^{\mi Ht}
\right],
\end{gather}
with a fixed environmental state $\rho_{\rm E}$ and a self-adjoint
Hamiltonian $H$ on
$\mathcal{H}_{\rm S}\otimes\mathcal{H}_{\rm E}$.
Suppose that the total Hamiltonian admits a decomposition into a free
non-interacting Hamiltonian $H_0=H_{\rm S}\otimes I_{\rm E}+I_{\rm S}\otimes H_{\rm E}$ and an interaction term $H_{\rm int}$, where $H_{\rm S}$ is a finite-dimensional self-adjoint operator on
$\mathcal{H}_{\rm S}$, $H_{\rm E}$ is a self-adjoint operator on
$\mathcal{H}_{\rm E}$, and $H_{\rm int}$ is a self-adjoint operator on
$\mathcal{H}_{\rm S}\otimes\mathcal{H}_{\rm E}$. 
Mathematically, this condition means that the operator $H_0+H_{\rm int}$, defined on
$D(H_0)\cap D(H_{\rm int})$, is assumed to be essentially self-adjoint,
and $H$ denotes its unique self-adjoint extension. In other words, $H$ is fixed without choosing an additional
self-adjoint extension when $H_0$ and $H_{\rm int}$ are given.

Under the mathematical assumption given above, we give a physical energy condition. For every pure state $\ket{\psi}$ of the system of interest,
there exists $\delta_\psi>0$ such that
\begin{gather}
C_\psi
:=
\sup_{0\leq s\leq\delta_\psi}
\Tr\!\left[
\rho_0(s)H_{\rm int}^2
\right]
<\infty ,
\label{eq:C-psi}
\end{gather}
where
\begin{gather}
\rho_0(0)
:=
\ketbra{\psi}{\psi}\otimes\rho_{\rm E},
\\
\rho_0(s)
:=
\me^{-\mi H_0 s}\rho_0(0) \me^{\mi H_0 s}.
\label{eq:rho-free}
\end{gather}
Then, for every pure state $|\psi\rangle$,
\begin{gather}
\ell(t)=O(t^2),
\end{gather}
and in particular
\begin{gather}
\lim_{t\to +0}
\frac{\ell(t)}{t}
=
0 .
\end{gather}
\end{theorem}

This condition means that the interaction has finite second moment along
the uncoupled short-time evolution.

Theorem 2 does not require
a finite total energy expectation with respect to $H$, but instead
assumes a finite short-time interaction strength along the free evolution
generated by $H_0$. In this case, the usual quadratic short-time
mechanism survives at the level of the reduced survival probability. In
analogy with
\begin{align}
&\abs{\braket{\phi(0)|\phi(t)}}^2\notag
\\
&=
1-
\left(
\braket{\phi(0)|H^2|\phi(0)}-\braket{\phi(0)|H|\phi(0)}^2
\right)t^2
+o(t^2),
\end{align}
the interaction regularity condition gives
\begin{gather}
\ell(t)=O(t^2).
\end{gather}

We next give the short-time behavior of the survival loss for GKLS
semigroups.

\begin{lemma}[Linear short-time loss for GKLS semigroups]
\label{lem:gksl-linear-loss}
On the finite-dimensional system, consider a time-independent GKLS generator
\begin{align}
\mathcal{L}(\rho)
=
-\mi[H_{\rm S},\rho]
+
\sum_k \gamma_k
\left(
L_k\rho L_k^\dagger
-
\frac{1}{2}
\left\{
L_k^\dagger L_k,\rho
\right\}
\right),
\end{align}
with $\gamma_k\geq0$, a Hermitian operator $H_{\rm S}$, and operators
$L_k$ on $\mathcal{H}_{\rm S}$.
For the semigroup $\mathcal{E}_t=\me^{t\mathcal{L}}$, the loss of the
survival probability from a pure state satisfies
\begin{align}
\ell^{\rm GKLS}(t)
&:=
1-
\Tr_{\rm S}
\left[
|\psi\rangle\langle\psi|
\mathcal{E}_t(|\psi\rangle\langle\psi|)
\right]
\\
&=
\Gamma t+O(t^2),
\end{align}
where
\begin{align}
\Gamma
=
\sum_k \gamma_k
\left(
\langle\psi|L_k^\dagger L_k|\psi\rangle
-
|\langle\psi|L_k|\psi\rangle|^2
\right)
\geq0,
\end{align}
see the Appendix for details of the proof.
The Hamiltonian part of $\mathcal{L}$ does not contribute to
$\Gamma$. Moreover, $\Gamma=0$ for all pure states
$|\psi\rangle$ if and only if the dissipative part of the generator
vanishes, equivalently the semigroup is unitary. Hence every non-unitary
GKLS semigroup has at least one pure state for which
\begin{gather}
\lim_{t\to +0}
\frac{\ell^{\rm GKLS}(t)}{t}
=
\Gamma
>
0 .
\end{gather}
Thus, a linear short-time loss appears in Markovian dissipation for at
least one pure initial state.
\end{lemma}

Combining Theorems~\ref{thm:spectral-regularity}
and~\ref{thm:interaction-regularity} with
Lemma~\ref{lem:gksl-linear-loss}, we see that both conditions exclude a
initial linear decay, while every non-unitary GKLS semigroup necessarily
has one for some pure state.
Consequently, under either regularity condition, the semigroup must be
unitary. Equivalently, a non-unitary GKLS semigroup cannot be
exactly realized under either condition.

\subsection{Proof of Theorem 1}
 We shift the Hamiltonian by
its lower bound and set $\bar{H}:=H-E_g\geq0$. This shift does not change
the reduced dynamics except for a global phase. Let
\begin{gather}
    U(t):=\me^{-\mi \bar{H}t} .
\end{gather}
Then, we can rewrite the survival loss as
\begin{align}
    \ell(t)
    &=
    {\rm Tr}
    \left[
        Q U(t)\rho_0(0) U(t)^\dagger
    \right]
    \\
    &=
    \norm{Q U(t)\rho_0(0)^{1/2}}_2^2
    \\
    &=
    \norm{Q\left(U(t)-I\right)\rho_0(0)^{1/2}}_2^2 .
\end{align}
Here we have used the following facts:
\begin{gather}
    Q \rho_0(0)=0,
    \qquad
    \rho_0(0) Q=0,
    \qquad
    Q\rho_0(0)^{1/2}=0 .
\end{gather}
Since $Q$ is a projection, multiplication by $Q$ is contractive in the
Hilbert--Schmidt norm. Thus we have
\begin{align}
    \ell(t)
    &\leq
    \norm{\left(U(t)-I\right)\rho_0(0)^{1/2}}_2^2 .
\label{eq:loss_bound}
\end{align}

Writing the spectral decomposition as
\begin{gather}
    \bar{H}=\int E\,\Pi(\dd E),
\end{gather}
we obtain the right-hand side of \eqref{eq:loss_bound} as
\begin{align}
    &\norm{\left(U(t)-I\right)\rho_0(0)^{1/2}}_2^2
    \notag\\
    &={\rm Tr}
    \left[
        \rho_0(0)^{1/2}
        \left(U(t)^\dagger-I\right)
        \left(U(t)-I\right)
        \rho_0(0)^{1/2}
    \right]
    \\
    &={\rm Tr}
    \left[
        \left(2I-U(t)-U(t)^\dagger \right)
        \rho_0(0)
    \right]
    \\
    &=\int {[2-2\cos(Et)]}\,{\rm Tr}\left[\Pi(\dd E)\rho_0(0)\right],
\end{align}
where we have used the fact that $U(t)-I$ is a bounded operator.

Therefore, we obtain
\begin{align}
    \frac{\ell(t)}{t}
    \leq
    \int
    \frac{2-2\cos(Et)}{t}\,
    \mu(\dd E),
\end{align}
where
\begin{gather}
    \mu(\dd E)
    :=
    {\rm Tr}
    \left[
        \Pi(\dd E)\rho_0(0)
    \right].
\end{gather}
Since $\bar{H}\geq0$, the support of $\mu$ lies in $E\geq0$. For each
fixed $E\geq0$,
\begin{gather}
    \lim_{t\to +0}
    \frac{2-2\cos(Et)}{t}
    =
    0 .
\end{gather}
Moreover, using $1-\cos x\leq x$ for $x\geq0$, we have
\begin{gather}
    0
    \leq
    \frac{2-2\cos(Et)}{t}
    \leq
    2E .
\end{gather}
The function $2E$ is integrable with respect to $\mu$ because
the finite-energy condition gives
\begin{align}
    \int E\,\mu(\dd E)
    &=
    {\rm Tr}
    \left[
        \rho_0(0) \bar{H}
    \right]
    <\infty .
\end{align}
Therefore, by the dominated convergence theorem,
\begin{gather}
    \lim_{t\to +0}
    \int
    \frac{2-2\cos(Et)}{t}\,
    \mu(\dd E)
    =
    0 .
\end{gather}
Since $\ell(t)\geq0$, this implies
\begin{gather}
    \lim_{t\to +0}
    \frac{\ell(t)}{t}
    =
    0 .
\end{gather}

\subsection{Proof of Theorem 2}
The survival loss is
\begin{align}
    \ell(t)
    &=
    \Tr\qty[
        Q\me^{-\mi Ht}\rho_0(0)\me^{\mi Ht}
    ]
    \\
    &=
    \norm{
        Q\me^{-\mi Ht}\rho_0(0)^{1/2}
    }_2^2 .
\label{eq:B-loss-HS}
\end{align}
 Let us introduce
\begin{gather}
    U_0(h)
    :=
    \me^{-\mi H_0h},
    \\
    V(h)
    :=
    \me^{-\mi H_{\rm int}h},
    \\
    K_n(t)
    :=
    \qty(U_0(h)V(h))^n,
\label{eq:B-Un-definitions}
\end{gather}
where $0<t\leq\delta_\psi$ and $h=t/n$.

By the assumed essential self-adjointness of $H_0+H_{\rm int}$ on
$D(H_0)\cap D(H_{\rm int})$, the Trotter--Kato product formula \cite{trotter1959product,kato1974trotter} gives:
\begin{equation}
    \me^{-\mi Ht}
    =
    \lim_{n\to\infty}
    \left(
        \me^{-\mi H_0t/n}
        \me^{-\mi H_{\rm int}t/n}
    \right)^n .
\label{eq:B-trotter-kato}
\end{equation}
Thus, the Trotter--Kato product formula gives the following strong convergence
\begin{gather}
    \lim_{n\to \infty} K_n(t)
    =
    \me^{-\mi Ht}.
\label{eq:B-strong-convergence}
\end{gather}
Since $\rho_0(0)^{1/2}$ belongs to the Hilbert--Schmidt class, and $Q$ and $K_n(t)$ are bounded operators, the strong convergence in \eqref{eq:B-strong-convergence}
yields convergence in the Hilbert--Schmidt norm:
\begin{gather}
    QK_n(t)\rho_0(0)^{1/2}\to Q\me^{-\mi Ht}\rho_0(0)^{1/2}.
\end{gather}
Therefore,
\begin{align}
    \ell(t)
    &=
    \lim_{n\to\infty}
    \ell_n(t),
    \\
    \ell_n(t)
    &:=
    \norm{
        QK_n(t)\rho_0(0)^{1/2}
    }_2^2 .
\label{eq:B-loss-trotter-limit}
\end{align}

We now split the approximate survival loss into a free part and an interaction part. Since
\begin{gather}
    K_n(t)
    =
    \me^{-\mi H_0t}
    +
    \qty(
        K_n(t)-\me^{-\mi H_0t}
    ),
\label{eq:B-loss-decomposition-operator}
\end{gather}
using
$\norm{A+B}_2^2\leq 2\norm{A}_2^2+2\norm{B}_2^2$, we have
\begin{align}
    \ell_n(t)
    &\leq
    2\ell_{\rm free}(t)
    +
    2r_n(t),
\label{eq:B-loss-decomposition}
\end{align}
where
\begin{gather}
    \ell_{\rm free}(t)
    :=
    \norm{
        Q\me^{-\mi H_0t}\rho_0(0)^{1/2}
    }_2^2,
    \\
    r_n(t)
    :=
    \norm{
        Q
        \qty(
            K_n(t)-\me^{-\mi H_0t}
        )
        \rho_0(0)^{1/2}
    }_2^2 .
\label{eq:B-free-and-remainder}
\end{gather}
Thus it is enough to show that both
$\ell_{\rm free}(t)$ and $r_n(t)$ are of order $t^2$, uniformly in
$n$.

First, we estimate the free part. Since $H_0$ is the uncoupled
Hamiltonian, the free unitary factorizes:
\begin{gather}
    \me^{-\mi H_0t}
    =
    \me^{-\mi H_{\rm S}t}
    \otimes
    \me^{-\mi H_{\rm E}t}.
\label{eq:B-free-factorization}
\end{gather}
Hence
\begin{align}
    \ell_{\rm free}(t)
    &=
    1-
    \Tr\qty[
        P\me^{-\mi H_0t}\rho_0(0)\me^{\mi H_0t}
    ]
   \\
    &=1-\Tr_{\mS}\qty[
        |\psi\rangle\langle\psi|
        \me^{-\mi H_{\rm S}t}
        |\psi\rangle\langle\psi|
        \me^{\mi H_{\rm S}t}
    ]\Tr_{\mE}\qty[
        \rho_{\rm E}
    ]\\
    &=
    1-
    \abs{
        \mel{\psi}{
        \me^{-\mi H_{\rm S}t}
        }{\psi}
    }^2 .
\label{eq:B-free-loss}
\end{align}
Since $\mathcal H_{\rm S}$ is finite-dimensional,
\begin{align}
 \ell_{\rm free}(t)
 &=\qty(\braket{\psi|H_{\rm S}^2|\psi}-\braket{\psi|H_{\rm S}|\psi}^2)t^2+o(t^2 ).
\label{eq:B-free-part-Ot2}
\end{align}

We next estimate the interaction part $r_n(t)$. 
We can easily check the following identity:
\begin{gather}
    A^n-B^n
    =
    \sum_{k=0}^{n-1}
    A^{n-1-k}(A-B)B^k.
\label{eq:B-telescopic-identity}
\end{gather}
Identifying $A=U_0(h)V(h)$ and $B=U_0(h)$, we have
\begin{align}
    &K_n(t)-\me^{-\mi H_0t}\notag
    \\&=
    \sum_{k=0}^{n-1}
    \qty(U_0(h)V(h))^{n-1-k}
    U_0(h)
    \qty(V(h)-I)
    U_0(kh).
\label{eq:B-telescopic-expansion}
\end{align}
Thus, the interaction part can be bounded as
\begin{align}
    &\norm{Q\qty(K_n(t)-\me^{-\mi H_0t})\rho_0(0)^{1/2}}_2\notag\\
    &=
    \Bigl\|
        Q\sum_{k=0}^{n-1}
        \qty(U_0(h)V(h))^{n-1-k}
        U_0(h)\notag
    \\
    &\qquad{}\times
        \qty(V(h)-I)
        U_0(kh)\rho_0(0)^{1/2}
    \Bigr\|_2\\
    &\leq
    \sum_{k=0}^{n-1}
    \Bigl\|
        \qty(U_0(h)V(h))^{n-1-k}
        U_0(h)
    \notag\\
    &\qquad{}\times
        \qty(V(h)-I)
        U_0(kh)\rho_0(0)^{1/2}
    \Bigr\|_2\\
    &=  \sum_{k=0}^{n-1}
        \norm{
            \qty(V(h)-I)
            U_0(kh)
            \rho_0(0)^{1/2}
        }_2
     .
\label{eq:B-remainder-direct-square}
\end{align}
Here we have used the invariance of the Hilbert--Schmidt norm under
multiplication by a unitary.
Therefore, we have
\begin{align}
    r_n(t)
    &\leq
    \qty(
        \sum_{k=0}^{n-1}
        \norm{
            \qty(V(h)-I)
            U_0(kh)
            \rho_0(0)^{1/2}
        }_2
    )^2 .\label{66}
\end{align}

Using
\begin{equation}
    \rho_0(s)^{1/2}
    =
    \me^{-\mi H_0s}\rho_0(0)^{1/2}\me^{\mi H_0s},
\label{eq:B-rho0-s-sqrt}
\end{equation}
 we can rewrite Eq.~\eqref{eq:C-psi} in the exact sense as
\begin{align}
   C_\psi &= \sup_{0\leq s\leq \delta_\psi} \Tr\qty[\rho_0(s)H_{\rm int}^2]
   \\
   &=
   \sup_{0\leq s\leq \delta_\psi}\norm{
      H_{\rm int}
      \me^{-\mi H_0 s}
      \rho_0(0)^{1/2}
   }_2^2<\infty .
\label{eq:B-trace-HS-equivalence}
\end{align}
This condition implies that
$H_{\rm int} U_0(kh)\rho_0(0)^{1/2}$ belongs to the Hilbert--Schmidt class for
$0\leq kh\leq t\leq\delta_\psi$. Let
\begin{gather}
    H_{\rm int}
    =
    \int_{\mathbb R}
        \lambda\, M(\dd\lambda)
\end{gather}
be the spectral decomposition of $H_{\rm int}$. We define a finite scalar
measure $\nu_{k,h}$ on $\mathbb R$ by
\begin{gather}
    \nu_{k,h}(\Delta)
    :=
    \norm{
        M(\Delta)
        U_0(kh)
        \rho_0(0)^{1/2}
    }_2^2,
\label{eq:B-nu-kh}
\end{gather}
where $\Delta$ is a Borel subset of $\mathbb R$. 
Then, using \eqref{eq:B-trace-HS-equivalence},
\begin{align}
    &\norm{
        \qty(V(h)-I)
        U_0(kh)
        \rho_0(0)^{1/2}
    }_2^2
    \notag\\
    &=
    \int_{\mathbb R}
        \abs{\me^{-\mi \lambda h}-1}^2\,
        \nu_{k,h}(\dd \lambda)
    \\
    &\leq
    h^2
    \int_{\mathbb R}
        \lambda^2\,
        \nu_{k,h}(\dd\lambda)
    \\
    &=
    h^2
    \norm{
        H_{\rm int}
        U_0(kh)
        \rho_0(0)^{1/2}
    }_2^2
    \leq
    h^2 C_\psi .
\label{eq:B-one-step-bound-squared}
\end{align}
Here we used
\begin{gather}
    \abs{\me^{-\mi \lambda h}-1}
    \leq
    |h\lambda|.
\label{eq:B-exp-bound}
\end{gather}
Equivalently,
\begin{gather}
    \norm{
        \qty(V(h)-I)
        U_0(kh)
        \rho_0(0)^{1/2}
    }_2
    \leq
    h\sqrt{C_\psi}.
\label{eq:B-one-step-bound}
\end{gather}
Substituting this into \eqref{66}, we obtain
\begin{align}
    r_n(t)
    &\leq
    \qty(
        \sum_{k=0}^{n-1}
        h\sqrt{C_\psi}
    )^2=
    C_\psi t^2 
\label{eq:B-remainder-Ot2}
\end{align}
because $h=t/n$.
This estimate is uniform in $n$.

Combining \eqref{eq:B-loss-decomposition},
\eqref{eq:B-free-part-Ot2}, and \eqref{eq:B-remainder-Ot2}, we find
\begin{align}
    \ell_n(t)
    &\leq
    2\ell_{\rm free}(t)
    +
    2C_\psi t^2=
    O(t^2),
\label{eq:B-elln-Ot2}
\end{align}
where the $O(t^2)$ bound is independent of $n$. Taking
$n\to\infty$ in \eqref{eq:B-loss-trotter-limit}, we conclude that
\begin{gather}
    \ell(t)
    =
    O(t^2)
    \qquad
    (t\to +0).
\label{eq:B-loss-Ot2}
\end{gather}
In particular,
\begin{gather}
    \lim_{t\to +0}
    \frac{\ell(t)}{t}
    =
    0 .
\label{eq:B-sublinear-limit}
\end{gather}

Thus, under the self-adjointness condition ensuring the Trotter--Kato product formula and the interaction regularity condition~\eqref{eq:C-psi}, the survival loss is quadratic at the initial time. Hence a
dissipative GKLS semigroup, which gives a linear initial loss for some
pure states, cannot be realized by such a regular time-independent
Hamiltonian dilation.

\section{Examples of Hamiltonians realizing exact GKLS equations}
\label{sec:examples}

In this section, we introduce two examples in which the reduced dynamics is exactly described by a GKLS semigroup. The first example is a model of a two-level system coupled to a bosonic bath that does not have a lower bound \cite{taira2024markovianity}. The reduced dynamics of this model is exactly described by the amplitude damping GKLS equation. In the second example, the total Hamiltonian is positive, but the reduced dynamics is exactly described by a dephasing GKLS equation \cite{burgarth2017positive}. While the Hamiltonian is positive, the energy expectation of the initial state is infinite, and the interaction term has an infinite second moment along the free evolution. 

The Hamiltonian of the first example is
\begin{gather}
    H
    =
    H_{\rm S}
    +
    H_{\rm E}
    +
    H_{\rm int},
    \label{eq:example-boson-H}
    \\
    H_{\rm S}
    =
    \frac{\omega_0}{2}\sigma_+\sigma_-,
    \qquad
    H_{\rm E}
    =
    \int_{-\infty}^{\infty}\dd k\,
    k\,b_k^\dagger b_k,
    \label{eq:example-boson-free}
    \\
    H_{\rm int}
    =
    g
    \int_{-\infty}^{\infty}\dd k\,
    \qty(
        \sigma_+ b_k
        +
        \sigma_- b_k^\dagger
    ).
    \label{eq:example-boson-int}
\end{gather}
Here $\sigma_+$ and $\sigma_-$ are the raising and lowering operators of
the two-level system, $b_k$ and $b_k^\dagger$ are the bosonic annihilation
and creation operators, and $g$ is a real constant. The reduced dynamics is exactly described by the amplitude damping GKLS equation.

This model violates the condition of Theorem 1 immediately. Since the bath dispersion
is $\omega(k)=k$ with $k\in\mathbb{R}$, the free environmental Hamiltonian
has no lower bound. Consequently, the total Hamiltonian is not bounded
from below.

The same model also violates the condition of Theorem 2. Let
\begin{gather}
    \rho_0(0)
    =
    \ketbra{1}{1}
    \otimes
    \ketbra{\rm vac}{\rm vac},
    \label{eq:example-boson-initial}
\end{gather}
where $\ket{1}$ is the excited state of the two-level system and
$\ket{\rm vac}$ is the bosonic vacuum. With
\begin{gather}
    H_0
    =
    H_{\rm S}
    +
    H_{\rm E},
    \label{eq:example-boson-H0}
\end{gather}
the state $\rho_0(0)$ is invariant under the free evolution up to a phase.
Hence
\begin{gather}
    \rho_0(s)
    =
    \me^{-\mi H_0s}
    \rho_0(0)
    \me^{\mi H_0s}
    =
    \rho_0(0) .
    \label{eq:example-boson-rho-free}
\end{gather}
Therefore,
\begin{align}
    \Tr\qty[
        \rho_0(0)H_{\rm int}^2
    ]
    &=
    \mel{1,{\rm vac}}{
        H_{\rm int}^2
    }{1,{\rm vac}}
    =
    g^2
    \int_{-\infty}^{\infty}\dd k,
    \label{eq:example-boson-Hint2}
\end{align}
and the right-hand side diverges. Thus, in the notation of Theorem 2,
\begin{gather}
    \sup_{0\leq s\leq\delta}
    \Tr\qty[
        \rho_0(s)H_{\rm int}^2
    ]
    =
    \infty
    \label{eq:example-boson-B-fails}
\end{gather}
for every $\delta>0$. Hence the exact Markovian behavior in this model is
not in conflict with the theorems. It is achieved precisely by using
a dilation outside the regular class.

We next consider an example showing that a lower bound of the Hamiltonian
alone is not sufficient. Let us consider the following positive Hamiltonian 
\begin{gather}
    H
    =
    \ketbra{0}{0}\otimes q_+
    +
    \ketbra{1}{1}\otimes q_- ,
    \label{eq:example-positive-H}
\end{gather}
where $q_\pm$ are multiplication operators on $L^2(\mathbb{R})$ defined by
\begin{gather}
    q_+\phi(x)
    =
    x_+\phi(x),
    ~~
    q_-\phi(x)
    =
    x_-\phi(x),
    \label{eq:example-positive-qpm}
    \\
    x_+
    =
    \max\{x,0\},
    ~~
    x_-
    =
    \max\{-x,0\}.
    \label{eq:example-positive-xpm}
\end{gather}
Since $q_\pm\geq0$, the Hamiltonian satisfies $H\geq0$.

Let us fix the environmental state as
\begin{gather}
    \rho_{\rm E}
    =
    \ketbra{\phi_C}{\phi_C},
\end{gather}
where
\begin{gather}
    \phi_C(x)
    =
    \braket{x|\phi_C}
    =
    \sqrt{\frac{\gamma}{2\pi}}\,
    \frac{1}{x-\omega_0-\mi\gamma/2}.
    \label{eq:example-positive-cauchy}
\end{gather}
Here $\gamma>0$ and $\omega_0\in\mathbb{R}$ are parameters.
In this case, the
reduced dynamics is described by a dephasing GKLS equation, although
\(H\geq0\).

To see how the regularity conditions of
Theorems~\ref{thm:spectral-regularity} and
\ref{thm:interaction-regularity} fail, we now choose the particular
initial product state
\begin{gather}
    \rho_0(0)
    =
    \ketbra{0}{0}
    \otimes
    \ketbra{\phi_C}{\phi_C}.
    \label{eq:example-positive-rho0}
\end{gather}
 The condition of Theorem 1 fails because the energy
expectation value is infinite:
\begin{align}
    \Tr\qty[
        \rho_0(0) H
    ]
    &=
    \mel{\phi_C}{
        q_+
    }{\phi_C}
    \\
    &=
    \int_0^\infty \dd x\,
    x\,|\phi_C(x)|^2
    \\
    &=
    \frac{\gamma}{2\pi}
    \int_0^\infty \dd x\,
    \frac{x}{
        (x-\omega_0)^2+\gamma^2/4
    } .
    \label{eq:example-positive-energy}
\end{align}
The integrand behaves as $1/x$ as $x\to\infty$, and hence the integral
diverges logarithmically:
\begin{gather}
    \Tr\qty[
        \rho_0(0) H
    ]
    =
    \infty .
    \label{eq:example-positive-energy-infinite}
\end{gather}
Thus, this example violates the condition of Theorem 1 not through the absence of a
lower bound, but through the heavy high-energy tail of the initial
environmental state.

Let us also examine the condition of Theorem 2. We decompose $H$ as
\begin{gather}
    H
    =
    H_0
    +
    H_{\rm int},
    \label{eq:example-positive-decomposition}
    \\
    H_0
    =
    \frac{1}{2}I_{\rm S}\otimes(q_++q_-),
    ~~
    H_{\rm int}
    =
    \frac{1}{2}\sigma_z\otimes(q_+-q_-).
    \label{eq:example-positive-H0-Hint}
\end{gather}
Since $q_++q_-=|q|$ and $q_+-q_-=q$, we have
\begin{gather}
    H_{\rm int}^2
    =
    \frac{1}{4}I_{\rm S}\otimes q^2,
    \label{eq:example-positive-Hint2}
\end{gather}
where $q$ is the position operator
\begin{gather}
    q\phi(x)
    =
    x\phi(x).
    \label{eq:example-positive-q}
\end{gather}
Moreover, $H_0$ and $H_{\rm int}$ commute. Therefore,
\begin{align}
    \Tr\qty[
        \rho_0(0)H_{\rm int}^2
    ]
    &=
    \mel{0,\phi_C}{
        H_{\rm int}^2
    }{0,\phi_C}
    \\
    &=
    \frac{1}{4}
    \mel{\phi_C}{
        q^2
    }{\phi_C}
    \\
    &=
    \frac{\gamma}{8\pi}
    \int_{-\infty}^{\infty}\dd x\,
    \frac{x^2}{
        (x-\omega_0)^2+\gamma^2/4
    } .
    \label{eq:example-positive-Hint-second}
\end{align}
The integrand approaches a nonzero constant as $|x|\to\infty$. Thus this
integral diverges and we have
\begin{gather}
    \sup_{0\leq s\leq\delta}
    \Tr\qty[
        \rho_0(s)H_{\rm int}^2
    ]
    =
    \infty
    \label{eq:example-positive-B-fails}
\end{gather}
for every $\delta>0$. Thus, the condition of Theorem 2 also fails.

These two examples show the role of the assumptions in the theorems. The
first example realizes exact Markovian behavior by using a Hamiltonian
with no lower spectral bound and an infinite interaction-energy moment.
The second example shows that even a positive Hamiltonian can generate
exact Markovian behavior only when the initial state or the interaction
sector is singular. In both cases, the obstruction is not the existence of
a unitary dilation itself, but the absence of a physically regular
Hamiltonian dilation in the sense of the conditions of Theorems 1 and 2.

\section{Conclusion}
We have investigated whether a dissipative time-independent
GKLS semigroup can be realized exactly as the reduced dynamics of a
physically regular time-independent Hamiltonian dilation. For a
finite-dimensional system, with the system initially prepared in a pure
state and the environment in a fixed state, we focused on the reduced
survival loss
\begin{gather}
    \ell(t)
    =
    1-
    \Tr_{\rm S}
    \left[
        \rho_{\rm S}(0)\rho_{\rm S}(t)
    \right].
\end{gather}
We proved that this quantity cannot have a positive linear initial decay
under either of two independent regularity assumptions on the microscopic
Hamiltonian dilation. Under the global regularity condition, namely a
lower-bounded total Hamiltonian with finite expectation in all product
states of a pure state of the system of interest and the fixed environmental state, the
survival loss satisfies
\begin{gather}
    \ell(t)=o(t).
\end{gather}
Under the interaction regularity condition for a decomposition
\(H=H_0+H_{\rm int}\), namely a finite short-time interaction-energy second
moment along the free evolution, it satisfies the stronger short-time
bound
\begin{gather}
    \ell(t)=O(t^2).
\end{gather}

This behavior is incompatible with a non-unitary GKLS semigroup. For any
time-independent GKLS generator with a non-vanishing dissipative part,
there exists at least one pure initial state for which
\begin{gather}
    \ell^{\rm GKLS}(t)
    =
    \Gamma t+O(t^2),
    \qquad
    \Gamma>0 .
\end{gather}
Thus, a non-unitary GKLS semigroup cannot be exactly obtained from a
time-independent Hamiltonian dilation satisfying either of the regularity
conditions considered here. Equivalently, if exact Markovian dissipation
is realized by a time-independent Hamiltonian dilation, then that dilation
must lie outside these regular classes.

The result may be viewed as an open-system counterpart of the
Chiu--Sudarshan--Misra short-time constraint. In the closed-system
setting, a lower-bounded Hamiltonian with finite initial energy forbids
linear short-time decay of the survival probability. Our result shows
that the same obstruction survives after tracing out an environment:
regular microscopic Hamiltonian dynamics cannot produce the strictly
linear decay that characterizes dissipative Markovian semigroups.

We also discuss the examples to illustrate how exact Markovian behavior
avoids this obstruction. In the
model giving the exact amplitude-damping GKLS equation, the environmental spectrum is unbounded from below and the
interaction-energy moment is already divergent. In
the positive-Hamiltonian dephasing model, the Hamiltonian is bounded from
below, but the initial environmental state has an infinite energy
expectation and the interaction-energy moment diverges. These examples
show that lower boundedness alone is not sufficient; finite energy
expectation and finite interaction-energy fluctuations are also essential
parts of physical regularity.

The obstruction identified here is therefore not a limitation on the
mathematical existence of unitary dilations. Rather, it concerns the
energy regularity of the Hamiltonian implementing such dilations. Exact
time-independent Markovian dissipation should be understood as a singular
energy-resource limit in which the universal short-time non-Markovian
regime has been removed. 

\begin{acknowledgments}
The author is deeply grateful to Hayato Kinkawa for helpful discussions and comments. 
The author also thanks Naomichi Hatano and Jaeha Lee for valuable comments. 
This work was supported by the RIKEN Junior Research Associate Program and the WINGS-QSTEP Program at the University of Tokyo.
  \end{acknowledgments}


\bibliography{apssamp}

@article{gorini1976completely,
  title={{Completely positive dynamical semigroups of N-level systems}},
  author={Gorini, Vittorio and Kossakowski, Andrzej and Sudarshan, Ennackal Chandy George},
  journal={Journal of Mathematical Physics},
  volume={17},
  number={5},
  pages={821--825},
  year={1976},
  publisher={American Institute of Physics}
}

@article{lindblad1976generators,
  title={{On the generators of quantum dynamical semigroups}},
  author={Lindblad, Goran},
  journal={Communications in mathematical physics},
  volume={48},
  number={2},
  pages={119--130},
  year={1976},
  publisher={Springer}
}

@article{davies1972some,
  title={{Some contraction semigroups in quantum probability}},
  author={Davies, EB},
  journal={Zeitschrift f{\"u}r Wahrscheinlichkeitstheorie und Verwandte Gebiete},
  volume={23},
  number={4},
  pages={261--273},
  year={1972},
  publisher={Springer}
}

@article{vom2023quantum,
  title={{Quantum-dynamical semigroups and the church of the larger Hilbert space}},
  author={vom Ende, Frederik},
  journal={Open Systems \& Information Dynamics},
  volume={30},
  number={01},
  pages={2350003},
  year={2023},
  publisher={World Scientific}
}

@article{taira2024markovianity,
  title={{Markovianity and non-Markovianity of Particle Bath with Dirac Dispersion Relation}},
  author={Taira, Takano and Hatano, Naomichi and Nishino, Akinori},
  journal={arXiv preprint arXiv:2406.17436},
  year={2024}
}

@article{burgarth2017positive,
  title={{Positive Hamiltonians can give purely exponential decay}},
  author={Burgarth, Daniel and Facchi, Paolo},
  journal={Physical Review A},
  volume={96},
  number={1},
  pages={010103},
  year={2017},
  publisher={APS}
}

@article{stinespring1955positive,
  title={{Positive functions on C*-algebras}},
  author={Stinespring, W Forrest},
  journal={Proceedings of the american mathematical society},
  volume={6},
  number={2},
  pages={211--216},
  year={1955},
  publisher={JSTOR}
}

@article{chiu1977time,
  title={{Time evolution of unstable quantum states and a resolution of Zeno's paradox}},
  author={Chiu, CB and Sudarshan, ECG and Misra, Baidyawath},
  journal={Physical Review D},
  volume={16},
  number={2},
  pages={520},
  year={1977},
  publisher={APS}
}

@book{breuer2002theory,
  title={{The theory of open quantum systems}},
  author={Breuer, Heinz-Peter and Petruccione, Francesco},
  year={2002},
  publisher={OUP Oxford}
}

@book{rivas2012open,
  title={{Open quantum systems}},
  author={Rivas, Angel and Huelga, Susana F},
  volume={10},
  year={2012},
  publisher={Springer}
}

@book{gardiner2004quantum,
  title={{Quantum noise: a handbook of Markovian and non-Markovian quantum stochastic methods with applications to quantum optics}},
  author={Gardiner, Crispin and Zoller, Peter},
  year={2004},
  publisher={Springer Science \& Business Media}
}

@article{jager2022lindblad,
  title={{Lindblad master equations for quantum systems coupled to dissipative bosonic modes}},
  author={J{\"a}ger, Simon B and Schmit, Tom and Morigi, Giovanna and Holland, Murray J and Betzholz, Ralf},
  journal={Physical Review Letters},
  volume={129},
  number={6},
  pages={063601},
  year={2022},
  publisher={APS}
}

@article{link2020dynamical,
  title={{Dynamical phase transitions in dissipative quantum dynamics with quantum optical realization}},
  author={Link, Valentin and Strunz, Walter T},
  journal={arXiv preprint arXiv:2005.10013},
  year={2020}
}

@article{prosen2008third,
  title={{Third quantization: a general method to solve master equations for quadratic open Fermi systems}},
  author={Prosen, Toma{\v{z}}},
  journal={New Journal of Physics},
  volume={10},
  number={4},
  pages={043026},
  year={2008}
}

@article{karevski2013exact,
  title={{Exact matrix product solution for the boundary-driven Lindblad XXZ chain}},
  author={Karevski, Dragi and Popkov, V and Sch{\"u}tz, GM},
  journal={Physical review letters},
  volume={110},
  number={4},
  pages={047201},
  year={2013},
  publisher={APS}
}

@article{ekman2024liouvillian,
  title={{Liouvillian skin effects and fragmented condensates in an integrable dissipative Bose-Hubbard model}},
  author={Ekman, Christopher and Bergholtz, Emil J},
  journal={Physical Review Research},
  volume={6},
  number={3},
  pages={L032067},
  year={2024},
  publisher={APS}
}

@article{verstraete2009quantum,
  title={{Quantum computation and quantum-state engineering driven by dissipation}},
  author={Verstraete, Frank and Wolf, Michael M and Ignacio Cirac, J},
  journal={Nature physics},
  volume={5},
  number={9},
  pages={633--636},
  year={2009},
  publisher={Nature Publishing Group UK London}
}

@article{schwartzman2025modeling,
  title={{Modeling error correction with Lindblad dynamics and approximate channels}},
  author={Schwartzman-Nowik, Zohar and Shirizly, Liran and Landa, Haggai},
  journal={Physical Review A},
  volume={111},
  number={2},
  pages={022613},
  year={2025},
  publisher={APS}
}

@article{malekakhlagh2025efficient,
  title={{Efficient Lindblad synthesis for noise model construction}},
  author={Malekakhlagh, Moein and Seif, Alireza and Puzzuoli, Daniel and Govia, Luke CG and van den Berg, Ewout},
  journal={npj Quantum Information},
  volume={11},
  number={1},
  pages={191},
  year={2025},
  publisher={Nature Publishing Group UK London}
}

@article{nathan2020universal,
  title={{Universal Lindblad equation for open quantum systems}},
  author={Nathan, Frederik and Rudner, Mark S},
  journal={Physical Review B},
  volume={102},
  number={11},
  pages={115109},
  year={2020},
  publisher={APS}
}

@article{mori2024liouvillian,
  title={{Liouvillian-gap analysis of open quantum many-body systems in the weak dissipation limit}},
  author={Mori, Takashi},
  journal={Physical Review B},
  volume={109},
  number={6},
  pages={064311},
  year={2024},
  publisher={APS}
}

@article{aydougan2025stabilizing,
  title={{Stabilizing steady-state properties of open quantum systems with parameter engineering}},
  author={Aydo{\u{g}}an, Koray and Schlimgen, Anthony W and Head-Marsden, Kade},
  journal={Physical Review Research},
  volume={7},
  number={2},
  pages={023057},
  year={2025},
  publisher={APS}
}

@article{leggett1987dynamics,
  title={{Dynamics of the dissipative two-state system}},
  author={Leggett, Anthony J and Chakravarty, SDAFMGA and Dorsey, Alan T and Fisher, Matthew PA and Garg, Anupam and Zwerger, Wilhelm},
  journal={Reviews of Modern Physics},
  volume={59},
  number={1},
  pages={1},
  year={1987},
  publisher={APS}
}

@article{hahn2025efficiency,
  title={{Efficiency of dynamical decoupling for (almost) any spin--boson model}},
  author={Hahn, Alexander and Burgarth, Daniel and Lonigro, Davide},
  journal={SciPost Physics},
  volume={19},
  number={2},
  pages={035},
  year={2025}
}

@article{trotter1959product,
  title={{On the product of semi-groups of operators}},
  author={Trotter, Hale F},
  journal={Proceedings of the American Mathematical Society},
  volume={10},
  number={4},
  pages={545--551},
  year={1959},
  publisher={JSTOR}
}

@article{kato1974trotter,
  title={{On the Trotter-Lie product formula}},
  author={Kato, Tosio},
  journal={Proceedings of the Japan Academy},
  volume={50},
  number={9},
  pages={694--698},
  year={1974},
  publisher={The Japan Academy}
}

\appendix*
\onecolumngrid
\section{Proof of the linear short-time loss for GKLS semigroups}
\label{app:gksl-linear-loss}

Let
\begin{gather}
    \rho_{\rm S}=\ketbra{\psi}{\psi}
\end{gather}
be a pure initial state. For the GKLS semigroup
\begin{gather}
    \rho_{\rm S}(t)=\me^{t\mathcal L}(\rho_{\rm S}),
\end{gather}
the survival loss is
\begin{align}
    \ell^{\rm GKLS}(t)
    &:=
    1-
    \Tr_{\rm S}\qty[
        \rho_{\rm S}\me^{t\mathcal L}(\rho_{\rm S})
    ]
    \\
    &=
    -\Tr_{\rm S}\qty[
        \rho_{\rm S}\mathcal L(\rho_{\rm S})
    ]t
    +O(t^2).
\end{align}
Substituting the GKLS form of \(\mathcal L\), we obtain
\begin{align}
    -\Tr_{\rm S}\qty[
        \rho_{\rm S}\mathcal L(\rho_{\rm S})
    ]
    &=
    -\Tr_{\rm S}\qty[
        \rho_{\rm S}
        \qty(
            -\mi[H_{\rm S},\rho_{\rm S}]
            +
            \sum_k\gamma_k
            \qty(
                L_k\rho_{\rm S}L_k^\dagger
                -
                \frac{1}{2}
                \qty{
                    L_k^\dagger L_k,\rho_{\rm S}
                }
            )
        )
    ]
    \\
    &=
    -\sum_k\gamma_k
    \Tr_{\rm S}\qty[
        \rho_{\rm S}
        \qty(
            L_k\rho_{\rm S}L_k^\dagger
            -
            \frac{1}{2}
            \qty{
                L_k^\dagger L_k,\rho_{\rm S}
            }
        )
    ],
\end{align}
because
\begin{gather}
    \Tr_{\rm S}\qty[
        \rho_{\rm S}[H_{\rm S},\rho_{\rm S}]
    ]
    =
    0.
\end{gather}
Using \(\rho_{\rm S}=\ketbra{\psi}{\psi}\), we find
\begin{align}
    -\Tr_{\rm S}\qty[
        \rho_{\rm S}\mathcal L(\rho_{\rm S})
    ]
    &=
    -\sum_k\gamma_k
    \qty[
        \abs{
            \bra{\psi}L_k\ket{\psi}
        }^2
        -
        \bra{\psi}L_k^\dagger L_k\ket{\psi}
    ]=:\Gamma.
\end{align}
Therefore,
\begin{gather}
    \ell^{\rm GKLS}(t)
    =
    \Gamma t+O(t^2).
\end{gather}
For each \(k\), the Cauchy--Schwarz inequality gives
\begin{gather}
    \bra{\psi}L_k^\dagger L_k\ket{\psi}
    =
    \norm{L_k\ket{\psi}}^2
    \geq
    \abs{
        \bra{\psi}L_k\ket{\psi}
    }^2.
\end{gather}
Thus \(\Gamma\geq 0\).

If \(\Gamma=0\) for every pure state \(\ket{\psi}\), then for every
\(k\) with \(\gamma_k>0\),
\begin{gather}
    L_k\ket{\psi}\propto\ket{\psi}
\end{gather}
holds for every \(\ket{\psi}\). Hence \(L_k\) is proportional to the
identity operator,
\begin{gather}
    L_k=c_k I_{\rm S}.
\end{gather}
In this case, the corresponding dissipative term vanishes:
\begin{gather}
    L_k\rho L_k^\dagger
    -
    \frac{1}{2}
    \qty{
        L_k^\dagger L_k,\rho
    }
    =
    0.
\end{gather}
Therefore, \(\Gamma=0\) for every pure state if and only if the
dissipative part of the GKLS generator vanishes. Equivalently, every
non-unitary GKLS semigroup has at least one pure state for which
\begin{gather}
    \lim_{t\to +0}
    \frac{\ell^{\rm GKLS}(t)}{t}
    =
    \Gamma
    >
    0.
\end{gather}
This proves the lemma.
\end{document}